\documentclass[aps,prx,longbibliography,a4paper,notitlepage,reprint,superscriptaddress]{revtex4-2}

\usepackage{amsmath,amssymb,amsfonts} 
\usepackage{mathrsfs} 

\usepackage{mathtools} 

\usepackage{color}
\usepackage{chemformula}
\usepackage{siunitx}
\usepackage{graphicx,float}
\usepackage[colorlinks,
    linkcolor=red,
    citecolor=blue,
    urlcolor=red]{hyperref}
\usepackage{blindtext}

\usepackage{cleveref}
\usepackage{soul} 
\usepackage{changes}
\usepackage{footmisc}

\newcommand{\SiN}{\ch{Si_3N_4}}

\newcommand{\figref}[1]{Fig.~\ref{#1}}
\renewcommand{\eqref}[1]{Eq.~\ref{#1}}

\begin{document}

\title{Strained crystalline nanomechanical resonators with ultralow dissipation}

\author{A. Beccari}
\affiliation{Institute of Physics, Swiss Federal Institute of Technology Lausanne (EPFL), 1015 Lausanne, Switzerland}

\author{D. A. Visani}
\affiliation{Institute of Physics, Swiss Federal Institute of Technology Lausanne (EPFL), 1015 Lausanne, Switzerland}

\author{S. A. Fedorov}
\affiliation{Institute of Physics, Swiss Federal Institute of Technology Lausanne (EPFL), 1015 Lausanne, Switzerland}

\author{M. J. Bereyhi}
\affiliation{Institute of Physics, Swiss Federal Institute of Technology Lausanne (EPFL), 1015 Lausanne, Switzerland}

\author{V. Boureau}
\affiliation{Interdisciplinary Center for Electron Microscopy (CIME), EPFL, 1015 Lausanne, Switzerland}

\author{N. J. Engelsen}
\email{nils.engelsen@epfl.ch}
\affiliation{Institute of Physics, Swiss Federal Institute of Technology Lausanne (EPFL), 1015 Lausanne, Switzerland}

\author{T. J. Kippenberg}
\email{tobias.kippenberg@epfl.ch}
\affiliation{Institute of Physics, Swiss Federal Institute of Technology Lausanne (EPFL), 1015 Lausanne, Switzerland}

\date{\today}
\maketitle
	

\noindent\textbf{In strained mechanical resonators, the concurrence of tensile stress and geometric nonlinearity dramatically reduces dissipation. This phenomenon, dissipation dilution \cite{gonzalez2000suspensions}, is employed in mirror suspensions of gravitational wave interferometers \cite{gonzalez1994brownian} and at the nanoscale, where soft-clamping \cite{tsaturyan2017ultracoherent} and  strain engineering \cite{ghadimi2018elastic} have allowed extremely high quality factors. However, these techniques have so far only been applied in amorphous materials, specifically \SiN. Crystalline materials exhibit significantly lower intrinsic damping at cryogenic temperatures, due to the absence of two level systems in the bulk, as exploited in Weber bars \cite{braginsky1985systems} and silicon optomechanical cavities \cite{maccabe2020nanoacoustic}. 
Applying dissipation dilution engineering to strained crystalline materials could therefore enable extremely low loss nanomechanical resonators, due to the combination of reduced internal friction, high intrinsic strain, and high yield strength \cite{sementilli2021nanomechanical}. Pioneering work \cite{liu2011highq,cole2014tensilestrained,kermany2014microresonators,romero2020engineering} has not yet fully exploited this potential. Here, we demonstrate that single crystal strained silicon, a material developed for high mobility transistors, can be used to realize mechanical resonators with ultralow dissipation. We observe that high aspect ratio ($>10^5$) strained silicon nanostrings support MHz mechanical modes with quality factors exceeding $10^{10}$ at \SI{7}{K}, a tenfold improvement over values reported in \SiN. At \SI{7}{K}, the thermal noise-limited force sensitivity is approximately \SI{45}{zN/Hz^{1/2}}---approaching that of carbon nanotubes \cite{debonis2018ultrasensitive}---and the heating rate is only 60 quanta-per-second. Our nanomechanical resonators exhibit lower dissipation than the most pristine macroscopic oscillators and their low mass makes them particularly promising for quantum sensing and transduction.}

\begin{figure}[!ht]
\centering
\includegraphics[width=\columnwidth]{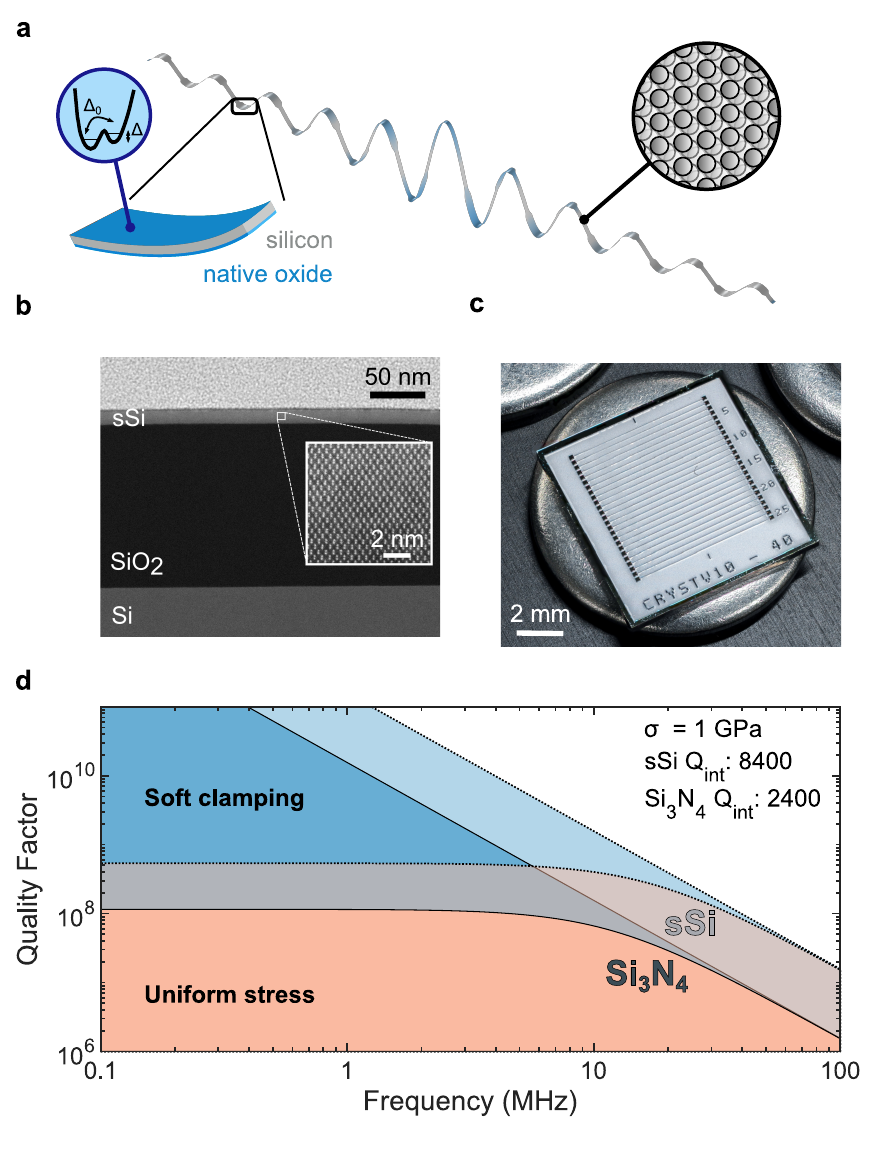}
\caption{\textbf{Dissipation in strained crystalline mechanical resonators.} \textbf{a}, Localized flexural mode of a strained silicon nanostring, simulated with finite element methods (FEM). 
\textbf{b}, Scanning transmission electron micrograph of the cross-section of a processed sSOI sample. The inset shows the strained silicon crystalline lattice as seen from a $\langle110\rangle$ direction. \textbf{c}, Photograph of a chip with an array of strained silicon strings. 
\textbf{d}, Diagram of dissipation dilution, with boundaries of colored regions indicating the Q vs. frequency limits for resonators with \SI{12}{nm} thickness and lengths below \SI{6}{mm}. 
\label{fig:overview}}
\end{figure}

\begin{figure*}[t]
\centering
\includegraphics[width=\textwidth]{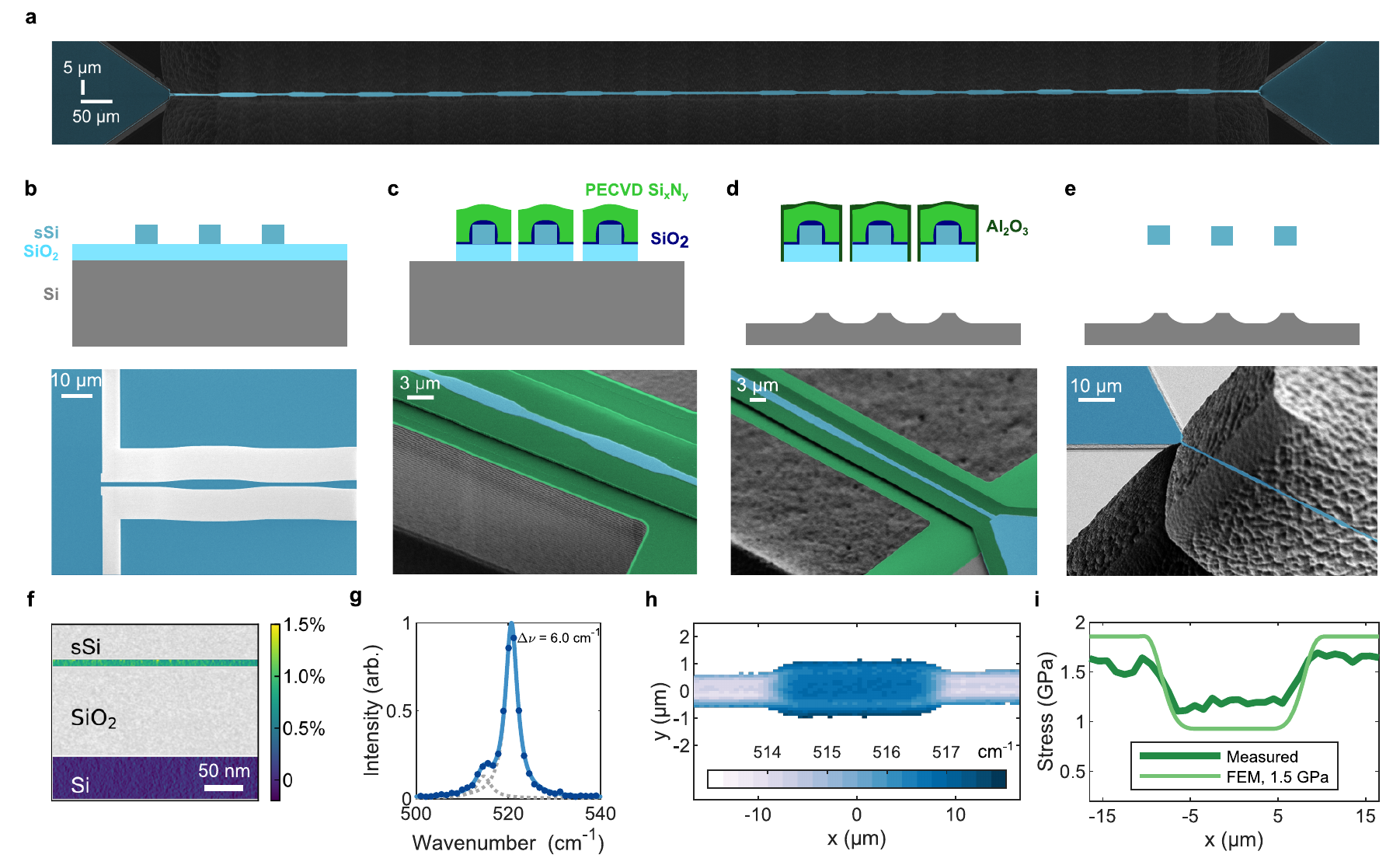}
\caption{\textbf{Fabrication of strained silicon nanostrings.} \textbf{a}, False-color SEM micrograph of a high aspect ratio PnC nanostring. Different scale factors have been applied in the horizontal and vertical directions. \textbf{b-e}, Simplified fabrication process, divided into: \textbf{b} device layer patterning, \textbf{c} encapsulation of the nanostrings, \textbf{d} undercut and \textbf{e} release and removal of the encapsulation layer. False-colour SEM micrographs are shown for each step of the process. \textbf{f}, Color-coded strain map of crystalline silicon, measured with dark-field electron holography \cite{boureau2019strain}. \textbf{g}, Raman Stokes spectrum collected from a string support.
\textbf{h}, Micro Raman map over the unit cell of a corrugated nanostring. The colour scale represents the central frequency of Raman emission. \textbf{i}, Dark green: stress profile along a $y = 0$ cut in the unit cell, reconstructed from h. Light green: output of a finite-element simulation with $\SI{1.5}{GPa}$ initial stress.
\label{fig:fabrication}}
\end{figure*}

\begin{figure*}[t]
\centering
\includegraphics[width=\textwidth]{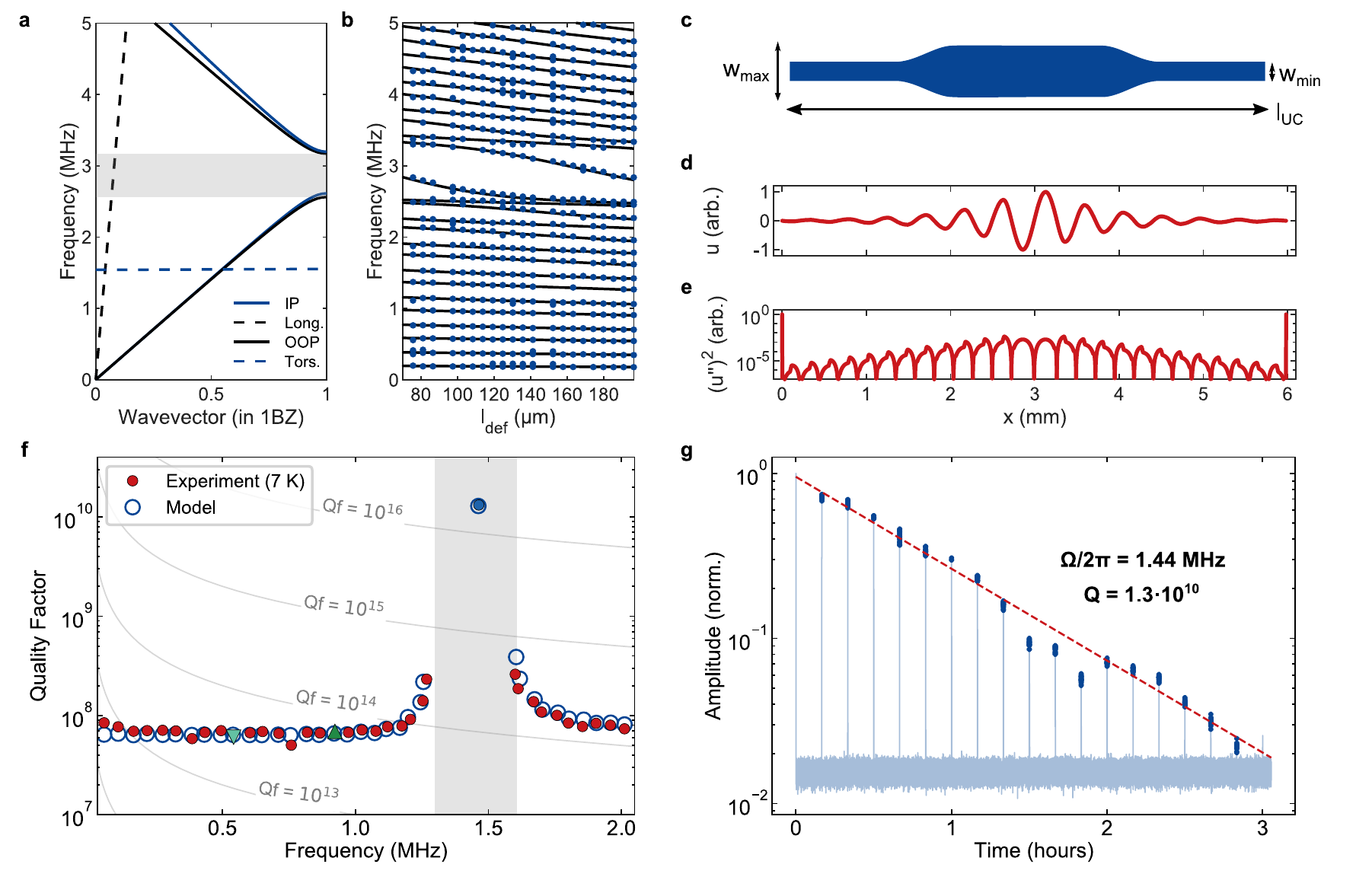}
\caption{\textbf{Bandgap engineering and soft clamping of nanostrings.} \textbf{a}, Vibrational bandstructure arising from the beam width corrugation. Branches: IP---in-plane flexural, Long.---longitudinal, OOP---out-of-plane, Tors.---torsional. The OOP bandgap is highlighted in gray. \textbf{b} OOP mode frequencies for strings with varying length of the center defect region. The length of the strings is about $\SI{1.6}{mm}$. \textbf{c}, Unit cell of the width corrugation profile. \textbf{d}, Displacement field and \textbf{e}, squared curvature of the localized mode of a $\SI{6.0}{mm}$-long nanostring, with 12 unit cells on each side of the defect. \textbf{f}, Quality factors and frequencies of the flexural modes of the $\SI{6.0}{mm}$-long nanostring, characterized at $\SI{7}{K}$. Measurement data are displayed with filled symbols, while predictions from a model supported by FEM simulations are displayed with blue empty circles. The mechanical bandgap region is shaded. The triangular and filled blue circle symbols mark modes analyzed in section \ref{sec:QvsT}. \textbf{g}, Gated ringdown of the localized mode in f (filled blue circle). An exponential fit is shown with a dashed red line. The small deviation from exponential decay around 1.5 hours may be due to alignment drifts in the characterization interferometer.
\label{fig:SC}}
\end{figure*}

Crystalline materials feature exceptionally low acoustic (mechanical) losses, as observed in early resonant bar gravitational wave detectors \cite{braginsky1985systems} and used for decades in stable frequency quartz oscillators \cite{cady1922piezoelectric}. The low internal dissipation of crystalline resonators, by virtue of the fluctuation-dissipation theorem, leads to lower thermomechanical noise, which motivated the transition from amorphous to crystalline materials as substrates for mirrors in interferometric gravitational wave detectors \cite{braginsky1999thermodynamical,hirose2014sapphire}. 
Furthermore, the development of crystalline mirror coatings \cite{cole2013tenfold}, substrates and spacers \cite{kessler2012sub40mhzlinewidth} have improved the frequency stability of reference cavities and lasers. Crystalline nanomechanical resonators have been employed in bolometric, inertial and magnetic force sensors \cite{junseokchae2005monolithic,rugar2004single,degen2009nanoscale} as well as for cavity optomechanical systems such as silicon optomechanical crystals \cite{safavi-naeini2013squeezed} and Fabry-P\'erot cavities with a movable mirror \cite{cripe2019measurement}. 

Amorphous or glassy materials, irrespective of their precise composition, exhibit universal properties due to the presence of two-level systems (TLSs), such as high dissipation at low temperatures \cite{phillips1987twolevel,pohl2002lowtemperature}. TLSs can couple to the strain field of acoustic vibrations \cite{grabovskij2012strain}, giving rise to acoustic absorption. In crystalline materials, TLSs only form due to defects and energy exchanges with TLS ensembles are drastically reduced. Through control of spurious damping channels, exceptionally low vibrational damping has been demonstrated in single-crystal resonators \cite{bagdasarov1974mechanical,galliou2013extremely,mcguigan1978measurementsa}, with quality factors $Q \sim 10^9\mathrm{-}10^{10}$.

Despite the high internal friction of amorphous materials, dissipation dilution has enabled extremely high quality factors for flexural modes of strained nanomechanical resonators \cite{unterreithmeier2010damping,yu2012control,fedorov2019generalized}. 
Dissipation dilution is characterized by the factor $D_Q$, which relates the $Q$ of a resonator mode to the intrinsic $Q$ of the material as $Q=D_Q\cdot Q_\mathrm{int}$. Here, $Q_\mathrm{int}= \phi^{-1}$, where $\phi$ is the phase delay between strain and stress, i.e. the material loss angle. A number of methods to engineer $D_Q$ have been developed in recent years \cite{tsaturyan2017ultracoherent,reetz2019analysis,bereyhi2019clamptapering,beccari2021hierarchical,ghadimi2018elastic,fedorov2019generalized}; with the majority of implementations reported in amorphous \SiN---a material that can be deposited with large tensile stress. Despite the relatively low $Q_\mathrm{int}$ of \SiN\ \cite{villanueva2014evidence}, quality factors up to $10^9$ were demonstrated at cryogenic temperatures, corresponding to a dilution factor ($D_Q$) exceeding $10^5$ \cite{rossi2018measurementbased,beccari2021hierarchical}. 

The combination of dissipation dilution and crystalline materials could enable even lower acoustic damping. In nanomechanical resonators, $Q_\mathrm{int}$ is known to be lower than in macroscopic devices, primarily due to pronounced surface effects such as friction in surface oxides, adventitious adsorbed layers or reconstructed surfaces \cite{braginsky1985systems,villanueva2014evidence, tao2015permanent}. Despite this, the use of crystalline materials can be highly advantageous: the temperature dependence of intrinsic mechanical dissipation processes often makes the coupling of phonons to TLSs the dominant loss mechanism at cryogenic temperatures. Several demonstrations of dissipation dilution in single crystal materials, such as \ch{GaAs} \cite{liu2011highq}, \ch{SiC} \cite{romero2020engineering,kermany2014microresonators} and \ch{InGaP} \cite{cole2014tensilestrained,buckle2018stress}, have been reported but could not attain lower dissipation than $\mathrm{Si_3N_4}$ devices. 

Here we implement strained silicon (sSi) mechanical resonators with ultralow dissipation. 
Strained silicon was developed as material for microelectronic devices, and has been used to improve the carrier mobility in MOSFETs \cite{chu2009strain} and to create a Pockels coefficient for optical modulation in silicon photonics \cite{jacobsen2006strained}, but nanomechanical applications of sSi remain unexplored thus far. We show that soft-clamped \cite{tsaturyan2017ultracoherent,ghadimi2018elastic}, strained silicon resonators (see \figref{fig:overview}a), achieve quality factors up to $1.3 \cdot 10^{10}$ for \SI{}{MHz} frequencies at temperatures $\sim \SI{7}{K}$. To the best of our knowledge, this is the highest Q reported for a mechanical oscillator at liquid \ch{He} temperatures, exceeded only at \SI{}{mK} temperatures in single-crystal silicon optomechanical cavities ($Q = 5\cdot 10^{10}$) \cite{maccabe2020nanoacoustic}.

\section{Strained silicon nanomechanics}

\label{sec:fab}

In this work, we use strained silicon on insulator (sSOI, Soitec SA) substrates, where the strained silicon layer is grown heteroepitaxially and bonded to a carrier wafer \cite{ghyselen2004engineering}. As in SOI technology, the sSi layer is conveniently separated from the silicon substrate by a buried oxide layer, facilitating the fabrication of suspended nanomechanical resonators. The average initial thicknesses of the sSi and buried oxide films in our samples are about \SI{14}{nm} and \SI{145}{nm}. A transmission electron microscope (TEM) cross-sectional image of the film stack is shown in \figref{fig:overview}b.

We developed a fabrication method to suspend high aspect ratio phononic crystal (PnC) nanostrings \cite{ghadimi2018elastic}, portrayed in \figref{fig:fabrication}a. Millimeter-scale nanostrings require a large clearance ($> \SI{10}{\micro\meter}$) from the substrate to be reliably suspended without stiction and collapse. An added complication is the chemical identity of the substrate and the device layer, offering no chemical selectivity for undercut steps. To circumvent these issues, our process includes the deposition of several encapsulation layers and a sequence of dry and wet etch steps for suspending the nanostrings (see \figref{fig:fabrication}b-e). The fabrication process is detailed further in the Appendices.

We assess the strain in the sSi film after fabrication by imaging a thin cross section of the chip, taken from the PnC nanostring pad, with a TEM. An interferometric electron optics technique, dark-field electron holography \cite{hytch2008nanoscale}, is employed to construct a strain map with nanometer resolution, displayed in \figref{fig:fabrication}f (see Appendices for further details). We measure an average biaxial strain of $\left(0.84 \pm 0.13\right)\%$ (stress $\left(1.51 \pm 0.23\right) \mathrm{GPa}$), in good agreement with the supplier specification (\SI{1.3}{GPa}). The strain is significantly higher than for stoichiometric \SiN\ on silicon \cite{villanueva2014evidence}, due to a lower (direction-averaged) Young's modulus.

Local strain can also be measured in suspended resonators by Raman spectroscopy \cite{dewolf1996stress}: for silicon, tensile (compressive) strain is known to decrease (increase) the Stokes scattering frequency. When collecting the Stokes signal from the sSOI stack, we observed spectra similar to the one depicted in \figref{fig:fabrication}g, where a red-shifted contribution from the sSi film can be resolved from the more intense peak at $\SI{521}{cm^{-1}}$, scattered by the unstrained substrate. We then collect Raman-scattered light from a suspended nanostring with nonuniform width and display the spatial variation of the Stokes frequency in \figref{fig:fabrication}h: the frequency changes along $x$, being lower in the thin parts of the unit cell. This occurs due to re-distribution of stress in a string with nonuniform width \cite{fedorov2019generalized}.
From the Stokes frequency we can extract the uniaxial stress along the unit cell (\figref{fig:fabrication}i), assuming previously-reported values for the phonon deformation potentials, and accounting for laser heating \cite{dewolf1996stress,suess2014powerdependent} (see Appendices). The reconstructed stress profile agrees fairly well with an initial stress of approximately $\SI{1.5}{GPa}$, as displayed in \figref{fig:fabrication}i. The imperfect quantitative matching of the profile may be due to an imprecise knowledge of the deformation potentials or small deviations of the elastic moduli from the bulk silicon values.

\section{Soft clamped PnC nanostrings}

\label{sec:SCresults}

In a strained string resonator, $D_Q$ for flexural modes takes the form \cite{fedorov2019generalized}:

\begin{equation}
    D_Q = \frac{1}{2\alpha_n\lambda+\pi^2\beta_n n^2\lambda^2},
    \label{eqn:DQ_string}
\end{equation}

\noindent where $n$ is the mode order, $\alpha_n$ and $\beta_n$ are factors which depend on the width profile of the string and on the mode shape, and $\lambda = \sqrt{\frac{1}{12\epsilon}}\frac{h}{L} \ll 1$, with $\epsilon$ the average static strain, $h$ the string thickness and $L$ the length. The dominant $\lambda$ term in the denominator of \eqref{eqn:DQ_string} stems from the displacement field curvature, $u''$, close to the clamping points of the string, while the $\lambda^2$ term originates from the curvature maxima associated with each antinode in the vibrational modeshape \cite{yu2012control}.

We apply the technique of soft clamping \cite{tsaturyan2017ultracoherent} to nanostrings using width corrugations, implementing a phononic crystal (PnC) and opening a bandgap around an acoustic wavelength twice as long as the width modulation period \cite{ghadimi2018elastic} (as portrayed in \figref{fig:SC}a). The resulting vibrational spectra are illustrated in \figref{fig:SC}b, displaying the out-of-plane, flexural modes (OOP) of soft-clamped nanostrings with $L \approx \SI{1.6}{mm}$. The width corrugation profile is illustrated schematically in \figref{fig:SC}c and is implemented with width modulation ${w_\mathrm{max}}/{w_\mathrm{min}}=2$ and unit cell length of $\approx \SI{108}{\micro\meter}$. A defect which perturbs the translational symmetry is implemented by stretching the length of the thin region in the string center ($l_\mathrm{def})$. The defect is surrounded on either side by 7 unit cells (see \figref{fig:fabrication}a). 

We characterize the devices in an interferometric characterization setup and acquire thermomechanical noise spectra to find the OOP frequencies of the resonators. To match the frequencies and the numerical predictions, we assume an initial stress in the sSi layer around $1.0 - \SI{1.2}{GPa}$, varying for different chips and fabrication runs. We speculate that this value, significantly lower than the stress reconstructed via TEM and Raman methods (see section \ref{sec:fab}), might be attributed to the presence of a thin layer of native oxide on the exposed silicon surfaces, which lowers the effective stress in the string cross-section \cite{tao2015permanent}. A bandgap manifests around $\SI{2.8}{}-\SI{3.2}{MHz}$. One or two resonances appear in the bandgap, corresponding to localized modes; their frequency responds much more sensitively to variations in $l_\mathrm{def}$ than those of distributed modes, as shown in \figref{fig:SC}b.
The displacement pattern of such a localized mode is displayed in \figref{fig:SC}d and exhibits soft clamping \cite{tsaturyan2017ultracoherent}, i.e. a suppressed clamping point curvature with respect to a uniform string (see \figref{fig:SC}e). For the localized mode, $\alpha_n \approx 0$ in \eqref{eqn:DQ_string}, and one is left with a more favorable $D_Q$ scaling:

\begin{equation}
    D_Q \approx \frac{1}{\pi^2\bar{n}^2\lambda^2} \propto \frac{\epsilon L^2}{h^2},
    \label{eqn:DQ_SC}
\end{equation}

\noindent where $\bar{n}$ is the index corresponding approximately to the center bandgap frequency. Soft clamped modes of high aspect ratio ($L/h$) strings exhibit an enhanced $D_Q$ compared to distributed modes, far from resonance with the PnC.

\begin{figure*}[!t]
\centering
\includegraphics[width=\textwidth]{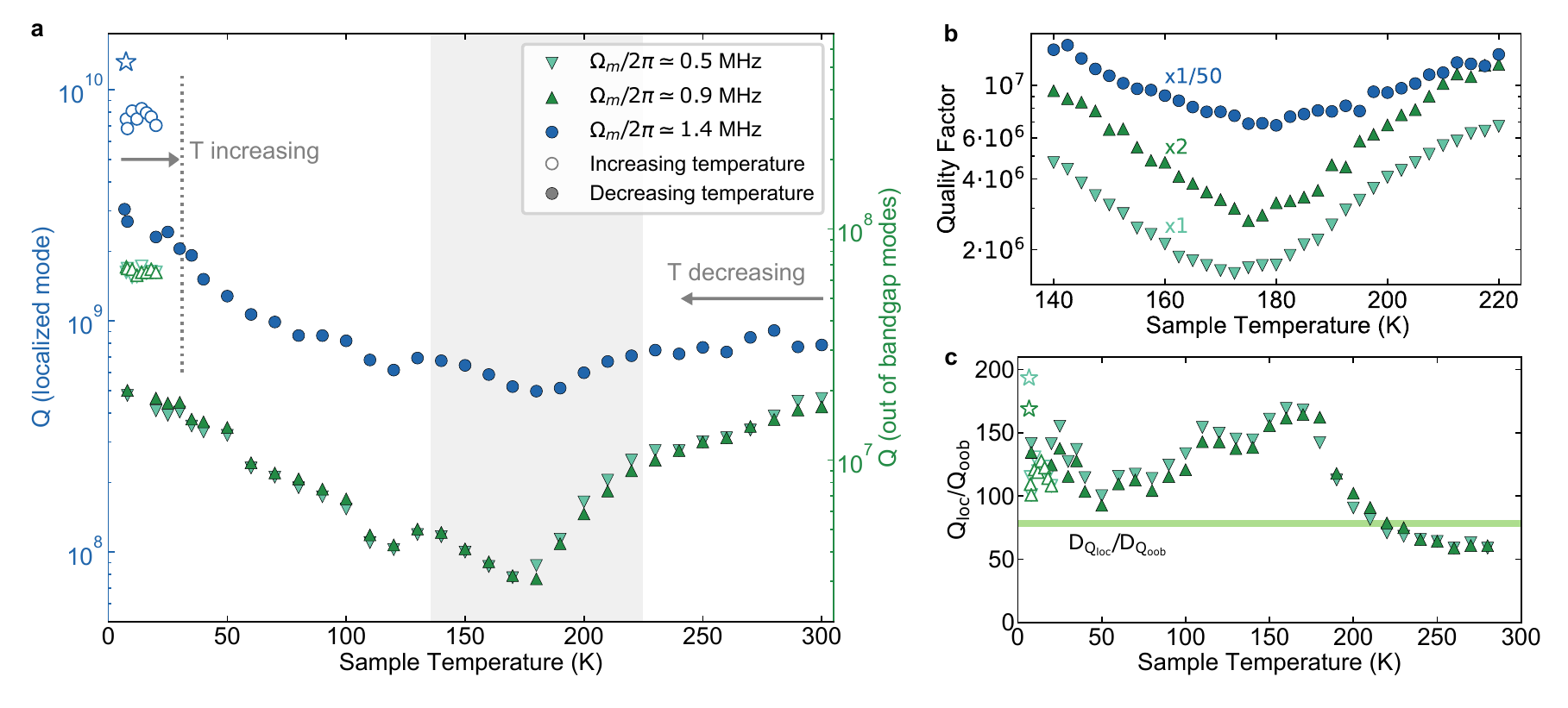}
\caption{\textbf{Temperature dependence of dissipation in a strained silicon nanostring.} \textbf{a}, Variation of $Q$ with temperature for different modes of a soft-clamped nanostring. Blue circles: localized mode, triangles: modes with frequencies below the mechanical bandgap. Empty symbols show measurements performed while increasing the temperature, closed symbols are measured later, as the temperature was decreased. Different axes are used for the quality factors of distributed modes (green) and the localized mode (blue). \textbf{b}, Finer temperature sweep around the region shaded in gray in a, where all modes present a dissipation peak. Scaling factors are applied to different series. \textbf{c}, Ratio of Qs between the localized mode and the two distributed modes, displayed with the same symbols as in a. The horizontal band displays an interval for the ratio of D\textsubscript{Q}s obtained with a finite element simulation.
\label{fig:QvsT}}
\end{figure*}

The enhanced $D_Q$ of soft clamped modes is observed in a $\SI{6.0}{mm}$-long nanostring, with 12 unit cells on each side of the defect ($l_\mathrm{def} \approx 2.3l_\mathrm{UC}$). The measurements are performed at the base temperature of our cryostat, $T\approx \SI{7}{K}$. The PnC localizes the mode with $n=26$ at $\Omega_m/2\pi = \SI{1.46}{MHz}$, shown in \figref{fig:SC}d. We evaluated the damping of OOP modes by initially driving them on then recording ringdown traces (see \figref{fig:SC}g). Since the laser beam increases the sample temperature (manifesting as a decrease of the resonant frequencies as optical power is increased), we perform gated detection by periodically blocking the probe laser with a shutter (see \figref{fig:SC}g). The repetition rate of the gates is much slower than thermal relaxation which occurs at timescales below \SI{1}{s}.

The measured $Q$s are shown in \figref{fig:SC}f: modes out of the vibrational bandgap display a weak dependence on frequency, due to the $\lambda^{-1}$ dependence seen in \eqref{eqn:DQ_string} for hard clamped modes. The mode $n=26$ (blue circle), localized to the defect region, exhibits a $Q \approx 1.3\cdot 10^{10}$, more than two orders of magnitude beyond the $Q$ of distributed modes. The $Q$ enhancement from soft clamping is too large to be explained by a single $Q_\mathrm{int}$ value; to match the measured points to the dissipation dilution predicted by a finite element simulation (open circles in \figref{fig:SC}f), we incorporated an additional loss term associated with the boundary regions in the model, affecting mostly the out-of-bandgap modes. This term can model phonon radiation to the substrate, or additional mechanical losses in the \ch{SiO2} layer underneath the clamping points. From the damping of the localized mode we estimate a $Q_\mathrm{int} \approx 8.1 \cdot 10^3$ for silicon at \SI{7}{K}, consistent with previous observations and surface-dominated mechanical losses \cite{tao2015permanent,yasumura2000quality}. 

\section{Temperature dependence of dissipation}

\label{sec:QvsT}

We then characterize the quality factors of several modes of the $\SI{6.0}{mm}$ nanostring as the sample holder temperature is tuned with a resistive heater. We probe two modes well below the lower bandgap edge ($\Omega_m/2\pi \sim \mathrm{500},\, \SI{900}{kHz}$) and the localized mode at $\Omega_m/2\pi \sim \SI{1.4}{MHz}$, highlighted in \figref{fig:SC}f. The results are shown in \figref{fig:QvsT}. We initially increase the sample temperature (measured with a thin film resistive thermometer connected to the sample plate) from \SI{7}{K} to \SI{20}{K} (open symbols in \figref{fig:QvsT}a), without significant variations of all the Qs. We then heat the sample to \SI{300}{K} and step down the temperature gradually (closed symbols in \figref{fig:QvsT}a). While the $Q$ of the localized mode around room temperature is partially limited by gas damping (see Appendices), we believe it to be negligible at low temperatures, due to lower pressures attained at cryogenic conditions ($< \SI[retain-unity-mantissa = false]{1e-8}{mbar}$).

Importantly, as the cryostat base temperature is reached again, we do not recover the initially observed Qs: the mechanical dissipation is increased by a factor of about 3. This observation corresponds to a gradual sample degradation over several days, which was reversible: we could reproduce the $Q > 10^{10}$ multiple times after heating the chip to room temperature and cooling it back to \SI{7}{K}. On the other hand, the mechanical frequencies drifted irreversibly to lower values, hinting to stress relaxation or to the growth of native oxide.

The Q of all modes has a minimum (the dissipation is peaked) around \SIrange{175}{180}{K}, and increases rapidly (but not monotonically) for lower temperatures. A closer examination of the relevant temperature range (\figref{fig:QvsT}b) shows that the temperature of the dissipation peak varies with the mode frequency, suggesting the presence of thermally-activated defects \cite{nowick1972anelastic,cannelli1991reorientation}, with activation energy around $\SI{0.2}{eV}$. The Q degradation is smaller for the localized mode, which could indicate a weaker coupling to the defect ensemble. A secondary dissipation peak is visible for all modes, around \SI{120}{K}. The absence of $Q$ peaks around \SI{125}{K} and \SI{20}{K}, where the linear expansion coefficient of silicon approaches zero excludes thermoelastic damping as a relevant dissipation source, as expected from the string dimensions \cite{gysin2004temperature}.

While the Qs of the two out-of-bandgap modes differ less than $15\%$ at any temperature, the ratio of Qs for localized and distributed modes exhibits a complex temperature dependence, shown in \figref{fig:QvsT}c. Were all modes affected by the same intrinsic damping mechanisms, the ratio would depend very weakly on the temperature and approximately correspond to the dilution factor ratio, $D_Q^\mathrm{loc}/D_Q^\mathrm{oob}$, displayed by the green band in \figref{fig:QvsT}c. The dependence of strain and string dimensions on temperature cannot explain the observed variation. As discussed in section \ref{sec:SCresults}, the PnC must therefore partially shield the localized mode from additional external loss channels, especially at cryogenic temperatures. At high temperatures, the $D_Q$ ratio approaches the numerical prediction; the small discrepancy can be explained by the contribution of gas damping to the dissipation of the localized mode at room temperature.

\section{Conclusions and outlook}

We have demonstrated strained silicon crystalline nanomechanical resonators with $Q > 10^{10}$, one order of magnitude beyond previous implementations in \SiN\ at liquid Helium temperatures \cite{rossi2018measurementbased, beccari2021hierarchical}. The built-in mechanical isolation of the soft clamped modes in this work eliminates many spurious loss mechanisms and provides access to the intrinsic material damping of silicon. 

The temperature scaling of dissipation could be extremely favourable at millikelvin temperatures, owing to low densities of TLSs. In light of recent demonstrations in unstrained optomechanical crystals of $Q_\mathrm{int} = 5\cdot10^{10}$ \cite{maccabe2020nanoacoustic}, soft clamped strained silicon devices might achieve $Q>10^{12}$, and thus probe the existence of novel mechanical dissipation mechanisms. The high yield stress of crystalline materials is particularly attractive for implementing strain engineered resonators \cite{sementilli2021nanomechanical,ghadimi2018elastic}, and the low intrinsic loss of silicon may be combined with other mode-shape engineering techniques \cite{bereyhi2019clamptapering,beccari2021hierarchical} to decrease damping even further. In addition, surface passivation techniques could be employed to diminish the effect of native oxide formation on dissipation \cite{tao2015permanent}.  At millikelvin temperatures, thermal transport in strings with low dimensionality might be particularly unconventional, and laser heating could be probed by examining the mechanical properties of the strings. The exceptionally high Q may also lend itself to force sensing, including recently-proposed searches for dark matter particles \cite{carney2021mechanicala}.

\section*{Appendices}

\subsection*{Detailed fabrication process}

Nanofabrication of the sSi mechanical resonators starts from \SI{300}{mm}-format sSOI wafers purchased from Soitec SA. In this fabrication technology, strained silicon is obtained by heteroepitaxial growth on \ch{Si_{1-x}Ge_{x}}; the crystal lattice constant mismatch can be tuned by changing the \ch{Ge} concentration $x$, leading to typical biaxial stress levels in excess of \SI{1}{GPa}. The epitaxially-strained silicon film is bonded to a silicon carrier wafer capped by an oxide layer, and separated from its original substrate by a sequence of ion implantation, thermal treatment and selective etch of the \ch{Si_{1-x}Ge_{x}} buffer layer \cite{ghyselen2004engineering}. Our wafers consist of a $\approx \SI{800}{\micro\meter}$-thick \ch{Si} substrate, with \SI{145}{nm} \ch{SiO2} and \SI{14}{nm} (nominal thickness) strained silicon thin films on the front side.

Wafers are resized through laser cutting to a \SI{100}{mm} diameter, suitable for handling and processing in our clean room. After thoroughly stripping the wafer of the protective layer of photoresist (with a combination of room temperature acetone and \ch{O2} plasma), we proceed to pattern the sSi layer. 

The electron beam resist \textsc{ZEP520A} at 50\% dilution is employed, spun at a thickness of roughly $\SI{150}{nm}$. During the electron beam exposure, the patterns are divided into a sleeve region, surrounding the nanostring edges and discretized with a fine grid, and a bulk region, including larger patches of the patterns with no common borders with the mechanical resonators. Proximity effect correction is performed to adjust the local exposure dose point by point. With this procedure, we can achieve line roughness $\lesssim \SI{10}{nm}$, while maintaining a reasonable duration for the exposure job. The nanobeams are patterned along $\langle 110 \rangle$ directions, where the stress relaxation upon release is negligible due to a minimal Poisson's ratio $\nu \approx 0.06$ (Young's modulus $E \approx \SI{169}{GPa}$). 

Using \textsc{ZEP520A} as a mask, we pattern the sSi layer with ICP-RIE, using \ch{SF6} + \ch{C4F8} gases. Only few seconds of exposure to plasma is needed to etch the film, landing on the \ch{SiO2} underneath. In order to increase reproducibility, we therefore pre-condition the ICP chamber by exposing a test \ch{Si} wafer to the same chemistry prior to etching the actual devices. The resist is stripped with a sequence of of hot N-Methyl-2-pyrrolidone (NMP) and \ch{O2} plasma; this procedure is carefully repeated after each etch step, to strip resist layers.

\begin{figure*}[t]
\centering
\includegraphics[width=\textwidth]{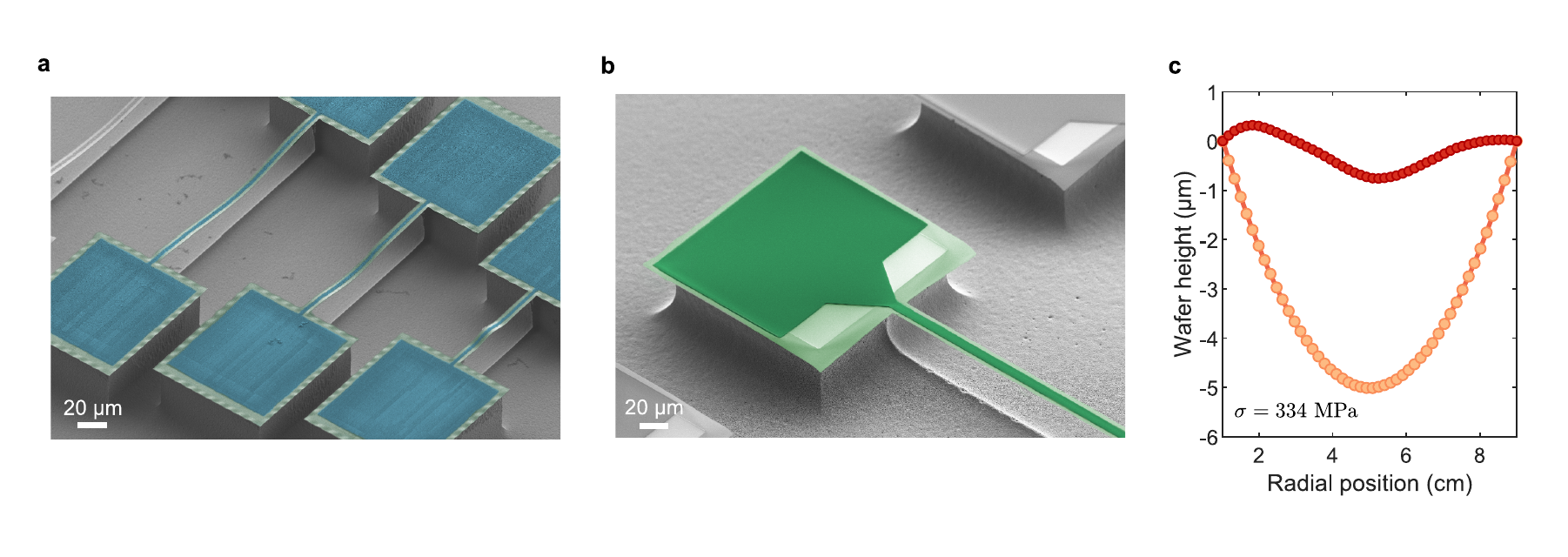}
\caption{\textbf{Fabrication of sSi nanostrings.} \textbf{a}, Suspended nanostrings without a stress-compensation layer are subject to a compressive load well beyond the critical value for out-of-plane buckling. Green - \ch{SiO2}, blue - patterned sSi + \textsc{FOX16} mask. \textbf{b}, The encapsulation layers (dark and light green) compensate compressive stress; the suspended beam is tensioned and flat. \textbf{c}, Determination of the intrinsic stress in PECVD \ch{Si_xN_y} through wafer bow. The wafer height profile is measured by scanning a laser beam and recording the angular deflection of the beam reflected off the surface. The magnitude and sign of the profile curvature permit to reconstruct the biaxial stress. Red circles---wafer profile before nitride deposition, orange circles---wafer bow after nitride deposition, with the initial profile subtracted.
\label{fig:suppfab}}
\end{figure*}

The goal of the process at this stage is to create sufficient clearance between the sSi layer and the substrate to suspend \SI{}{mm}-scale devices. It therefore seems appropriate to encapsulate the sSi strings in a protective oxide layer, then use deep reactive ion etching (DRIE) and an isotropic undercut (in a sequence similar to the method described in \cite{ghadimi2018elastic}) with high oxide chemical selectivity. However, an additional constraint arises: when such a multi-layer beam is suspended, it will buckle and very commonly collapse or break, due to the high compressive stress intrinsic to thermal and deposition oxides (see \figref{fig:suppfab}a). The solution is to ensure the encapsulating layer has tensile stress and is sufficiently thick to compensate the compressive stress in the buried oxide underneath (the presence of the sSi layer can be ignored, due to its negligible thickness), as in \figref{fig:suppfab}b.

In practice, multiple dielectric films are deposited. First, roughly $\SI{10}{nm}$ of \ch{SiO2} are deposited through atomic layer deposition (ALD), covering the patterned surface. Then, hydrogen-rich \ch{Si_xN_y} is grown through plasma-enhanced chemical vapour deposition (PECVD). With $\SI{40}{W}$ of RF power generating the plasma, the film exhibits a strong tensile stress, stable in time, of $\approx +\SI{330}{MPa}$ (characterized through the Stoney wafer bending method \cite{flinn1987measurement}; see \figref{fig:suppfab}c). Its thickness must be sufficient to keep the thickness-averaged stress positive, i.e. 

\begin{equation}
    t_\mathrm{SiN} > \frac{\sigma_\mathrm{SiO2} \left(1-\nu_\mathrm{SiO2}\right)}{\sigma_\mathrm{SiN} \left(1-\nu_\mathrm{SiN}\right)} t_\mathrm{SiO2},
\end{equation}

\noindent where $t$ is the film thickness and $\sigma, \nu$ are the initial stress ($\sigma_{\ch{SiO2}} \approx - \SI{360}{MPa}$) and Poisson's ratio. We usually deposit an approximately $\SI{400}{nm}$-thick nitride layer, verifying the thickness by reflectance spectroscopy.

This stack of dielectric films is patterned through conventional photolithography and RIE (\ch{He} + \ch{H2} + \ch{C4F8}), with a rescaled mask which completely envelopes the nanobeam sidewalls, exposing the \ch{Si} substrate. To improve the resilience of the encapsulation layer to the \ch{Si} undercut process, an additional $\SI{30}{nm}$ film of \ch{Al2O3} is deposited by ALD. Another photolithography mask defines the pattern of \ch{Al2O3} and protects the encapsulation structure during a DRIE step (Bosch process, pulsed \ch{SF6} and \ch{C4F8}), which creates a (roughly) $\SI{30}{\micro\meter}$-thick recess into the \ch{Si} substrate.

A thick, protective layer of photoresist i~s then spun over the patterned frontside, and the wafer is diced into chips; the process then continues chip-wise. Samples are cleaned carefully with a procedure similar to each post-RIE cleaning (taking care not to use Pira\~{n}a solution for stripping organic residue: it quickly dissolves \ch{Al2O3}). Selective undercut (\figref{fig:fabrication}d) is carried out with isotropic etching in \ch{SF6} plasma; for this step the single chips are bonded to a \ch{Si} carrier wafer in order to homogenize the etch rate and increase process reproducibility. Importantly, the separate patterning of the \ch{Al2O3} mask allows to control finely the geometry of the isotropic undercut process, preventing the formation of overhang at the string anchoring pads.

The strings are now suspended and encapsulated in a rescaled, thicker dielectric beam. To conclude the process and expose sSi one must remove the encapsulation layers selectively to \ch{Si}. For this purpose we employ highly concentrated \ch{HF} (25\% volume fraction in water solution). A non-negligible attack of the sSi layer in buffered HF (BHF) was, in fact, observed, possibly due to the higher pH of the etchant \cite{chen2005etching}. To ensure the survival of the fragile suspended samples during wet etch, it was crucial to design and construct appropriate PTFE chip holders that minimize the flow of liquids in the vicinity of the devices and allow gradual dilution of etchants and solvents, inspired by the \textit{turbulence-shielding} methods detailed in \cite{norte2015nanofabrication}. After the wet etch step, we transfer the samples to an ethanol bath and start the process of critical point drying (CPD) which removes the liquids while minimizing stiction forces.

The suspended strings can be exposed to \ch{O2} plasma for further cleaning of organic residues. Before the chips are loaded in the cryostat vacuum chamber, they are briefly exposed to vapor \ch{HF} for a few seconds to strip the spontaneously-formed native oxide layer and to improve the chemical stability of surfaces through hydrogen termination \cite{borselli2006measuring}. After vapor \ch{HF} exposure, the samples are brought to the cryostat vacuum chamber within 20 minutes.

\subsection*{Transverse buckling of unit cells}

\begin{figure}[t]
\centering
\includegraphics[width=\columnwidth]{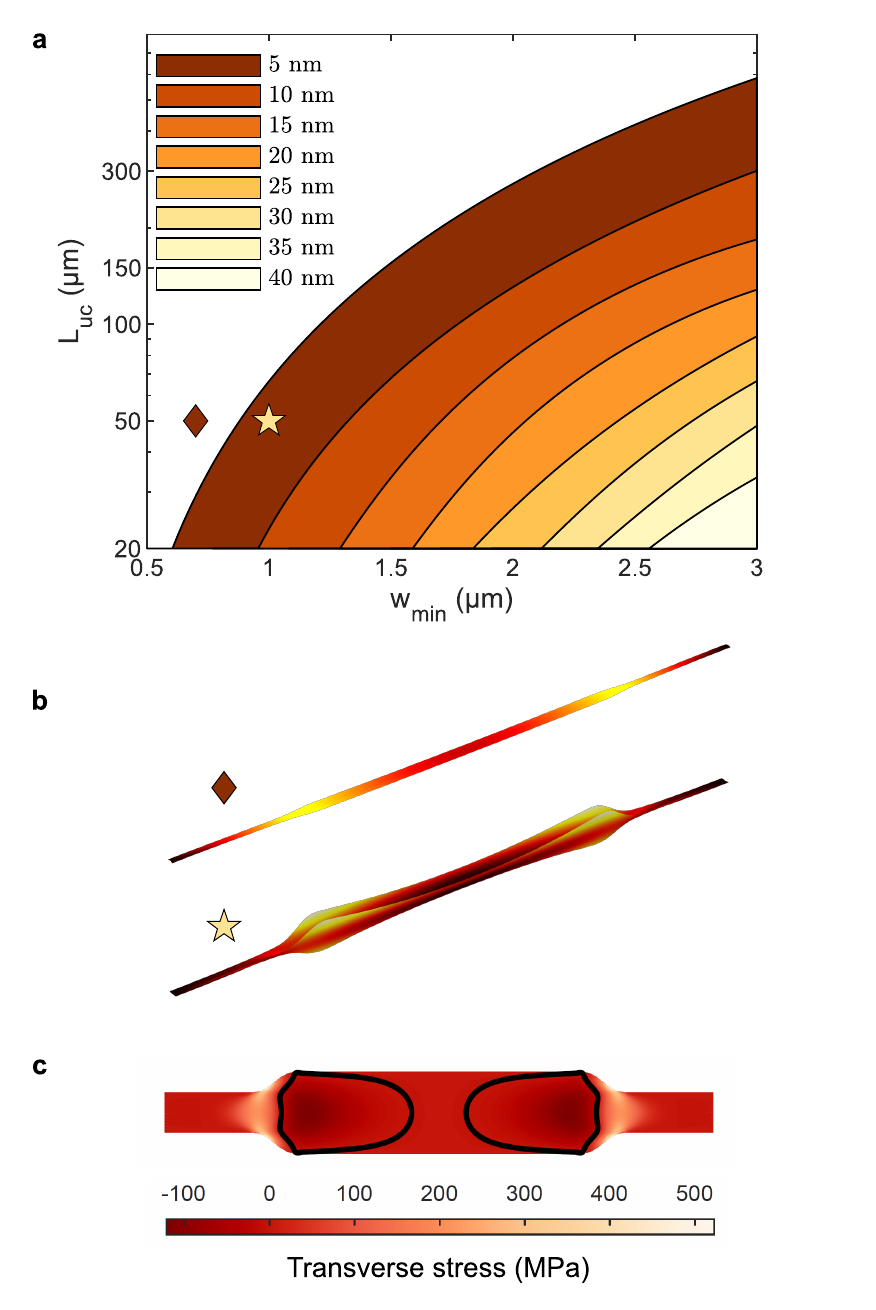}
\caption{\textbf{Conditions for transverse buckling.} \textbf{a}, Buckling instability contours for unit cells of varying thickness and in-plane dimensions. Coloured areas mark instability regions for fixed film thickness. \textbf{b}, Post-buckling deformation for a unit cell thickness of \SI{5}{nm} and in-plane dimensions marked with the corresponding symbol in a. \textbf{c}, Stress distribution in the transverse direction in a \SI{20}{\micro \meter}-long, \SI{1}{\micro \meter}-wide unit cell. The black lines depict the contours of zero transverse stress.
\label{fig:buckl}}
\end{figure}

PnC nanostrings are prone to localized buckling around the transition region of the width corrugation. This effect is due to the development of compressive stress in the direction transverse to the string axis, when the string is suspended and the biaxial stress is relaxed \cite{bereyhi2019clamptapering}.

This phenomenon is particularly relevant for very thin films, and occurs when the aspect ratio of the unit cell is decreased, i.e. for short and wide unit cells. 

In order to design PnC strings that do not exhibit buckling, we performed a finite element simulation of a single unit cell, with periodic boundary conditions. An eigenmode analysis is carried out, and the presence of eigenfrequencies with nonzero imaginary parts is interpreted as a manifestation of static instabilities leading to buckling. The simulation is repeated for different string aspect ratios, and the results are displayed in \figref{fig:buckl}a: coloured regions represent geometrical domains of instability for variable film thickness $h$. The area of instability regions increases as the thickness is decreased. The unit cells aspect ratios of all devices presented in the main text are chosen outside of the buckling region for $h = \SI{15}{nm}$. The simulation method can be validated by performing a stationary-state stress relaxation analysis with a small out-of-plane load applied to the unit cell surface, breaking the symmetry of the model: when the cell aspect ratio lies inside the coloured regions, the perturbation reveals the unstable character of the non-buckled solution and produces a visualization of the buckling mode of the unit cell (see \figref{fig:buckl}b).

Experimentally, the occurrence of buckling provides an upper bound on the mechanical frequency of modes that can be localized with high Q, especially when the stress and, correspondingly, the speed of sound are enhanced by width tailoring \cite{ghadimi2018elastic}.

\subsection*{Estimating the stress with electron holography}

A thin cross-sectional slice ($\sim \SI{100}{nm}$ thickness) is carved from the support pads of the nanobeams, using a focused ion beam (FIB) specimen preparation technique. The slice is observed in a transmission electron microscope (TEM) (see \figref{fig:overview}b). An electronic interference method, dark-field electron holography \cite{hytch2008nanoscale}, is employed to quantitatively map the strain in the crystalline films (Si substrate and bonded sSOI layer), with nanometer-scale resolution. 

The impinging electron wave is diffracted by the crystal of the specimen, set in two-beam condition to enhance the intensity of the (220)-diffracted beam which is selected with an objective aperture placed in the back-focal-plane of the objective lens of the TEM. Then, an electron biprism is used to interfere the reference wave emerging from the unstrained Si substrate (the dark-field image of the substrate) with the wave emerging from the measurement area (the dark-field image of the sSOI), to generate a hologram. Fourier processing is used to retrieve the geometric phase encoded in the interference fringes of the hologram, and the in-plane strain map is calculated from the gradient of the geometric phase, being proportional to the displacement field of the (220) crystal lattice planes \cite{boureau2019strain}

Electron holography experiments were performed with a double aberration-corrected FEI Titan Themis operated at \SI{300}{kV}. Stacks of 50 holograms of \SI{5}{s} exposure time were recorded with a FEI Ceta camera to improve the signal-to-noise ratio of the strain maps \cite{boureau2019strain}. The biprism bias was \SI{200}{V}, giving a hologram carrier frequency of \SI{0.69}{nm^{-1}}. The numerical aperture used for the Fourier processing limit the spatial resolution of the strain map to \SI{2.75}{nm}. 
.

\subsection*{Estimating the stress from Raman shifts}

Micro-Raman spectroscopy has been long recognized as a powerful technique to characterize the local stress state of crystalline samples, owing to the high spatial resolution achievable with confocal microscopy, and the high resolving power of diffraction gratings, compared to typical stress-induced Raman shifts \cite{dewolf1996stress}.

Silicon is arguably the best-studied material in this respect. The frequencies of long-wavelength optical phonons are perturbed by strain. In the absence of strain $\epsilon$, the normal modes are degenerate at a wavenumber of $\nu_0 \approx \SI{520}{cm^{-1}}$ ($\omega_0 = 2\pi\cdot \SI{15.6}{THz}$); the introduction of strain modifies the crystal stiffness tensor and perturbs the mode frequencies, breaking their degeneracy \cite{dewolf1996stress}. This influence is generally approximated as a linear dependence; by symmetries of the silicon lattice, only three independent elements appear in the stiffness-strain tensor, $p,q$ and $r$. These are usually determined experimentally, and different values have been reported in the literature \cite{suess2014powerdependent}. We employ here the numerical values from \cite{cerdeira1972stressinduced}:

\begin{align}
     p &= -1.39 \cdot \nu_0^2 \nonumber \\
     q &= -2.01 \cdot \nu_0^2 \\
     r &= -0.65 \cdot \nu_0^2 \nonumber
     \label{eqn:stiffcoeffs}
\end{align}

 The perturbed normal modes frequencies are obtained by finding the eigenvalues $\lambda$ of the matrix:
 
 \begin{footnotesize}
 
\begin{equation}
\begin{pmatrix}
p\epsilon_\mathrm{11} + q\left(\epsilon_\mathrm{22}+\epsilon_\mathrm{33}\right) &
2r\epsilon_\mathrm{12} &
2r\epsilon_\mathrm{13} \\
2r\epsilon_\mathrm{12} &
p\epsilon_\mathrm{22} + q\left(\epsilon_\mathrm{11}+\epsilon_\mathrm{33}\right) &
2r\epsilon_\mathrm{23} \\
2r\epsilon_\mathrm{13} &
2r\epsilon_\mathrm{23} &
p\epsilon_\mathrm{33} + q\left(\epsilon_\mathrm{11}+\epsilon_\mathrm{22}\right)
\end{pmatrix},
\label{eqn:secular}
\end{equation}

 \end{footnotesize}

\noindent from which the phonon wavenumbers can be computed as $\nu \approx \nu_0 + {\lambda}/{2\nu_0}$. Phonon polarizations are found as the corresponding eigenvectors. Raman scattering selection rules impose that only scattering by longitudinal phonons (polarized along z) can be measured, upon reflection on a $(1 0 0)$ surface \cite{ossikovski2008theory}. 

A particular strain state is assumed to simplify the eigenvalue equation and extract a single strain or stress component from the measurement of $\Delta\nu$. Importantly, matrix \ref{eqn:secular} is expressed in the reference system $[1 0 0], [0 1 0], [0 0 1]$, so that the strain tensor must be rotated to the same axes.

We report here the resulting linear relations between stress and Raman shift in the relevant cases.

\begin{itemize}
    \item Uniaxial stress directed along $[1 1 0]$ or $[1 0 0]$: $\Delta\nu/\sigma \approx \SI{-2.1}{cm^{-1}\per\giga\pascal}$
    \item Biaxial stress, $\sigma_\mathrm{xx} = \sigma_\mathrm{yy} = \sigma$: $\Delta\nu/\sigma \approx \SI{-4.3}{cm^{-1}\per\giga\pascal}$,
\end{itemize}

\noindent confirming the general observation that tensile (compressive) stress leads to a red (blue) shift in the Raman scattering peak.

In our confocal microscope setup, we excite Raman scattering by focusing a \SI{488}{nm} laser with typical output power of $\SI{10}{}-\SI{50}{mW}$ on the sample surface, we detect the Stokes peak in reflection and separate it from the laser line with a diffraction grating with high resolving power. The intensity and wavenumber of the Stokes peak are recorded by a CCD sensor. When the laser beam is focused on the suspended string, the spectrum consists of a single Lorentzian, since the recessed substrate is beyond the depth of focus of the microscope. On the other hand, when the laser is focused on a region where sSi is not undercut, two contributions can be distinguished: an intense Lorentzian line at $\nu_0$ collected from the unstrained substrate and a redshifted Lorentzian peak from the biaxially-stressed sSi layer. By extracting $\Delta\nu = \nu - \nu_0$ from the fit, the stress magnitude can be inferred as described above. We measure a Stokes shift from the unstrained substrate of $\nu_0 = \SI{521}{cm^{-1}}$, at small impinging optical power.

\begin{figure}[t!]
\centering
\includegraphics[width=\columnwidth]{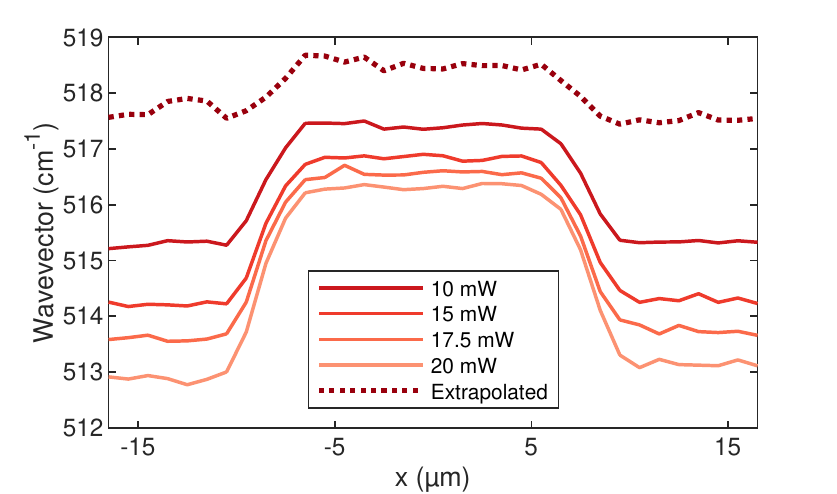}
\caption{\textbf{Raman scans with variable laser power.} Micro-Raman scans along a PnC beam unit cell are conducted, with variable laser power. A linear dependence of the Raman scattering wavevector on the power is observed; the dotted line indicates the extrapolation to zero optical power.
\label{fig:suppRaman}}
\end{figure}

Heating effects in the sSi layer influence the reconstructed stress magnitude. Silicon absorbs strongly at the pump laser wavelength, and heating is exacerbated by the poor thermal conductance of the nanostrings. A temperature increase leads to a redshift of the phonon frequency, due to anharmonic terms in the potential energy of atomic bonds \cite{hart1970temperature}, and changes the local strain. We account for this temperature dependence experimentally, by varying the pump power and extracting, point by point, the Stokes shift at vanishingly small power via linear extrapolation (see \figref{fig:suppRaman}). Note that the sensitivity of the Stokes frequency with respect to the impinging power in \figref{fig:suppRaman} varies with the laser position on the unit cell; this can be explained by the nonuniform thermal conductance of a PnC corrugated string.

\subsection*{Mechanical characterization setup}

\begin{figure*}[t]
\centering
\includegraphics[width=\textwidth]{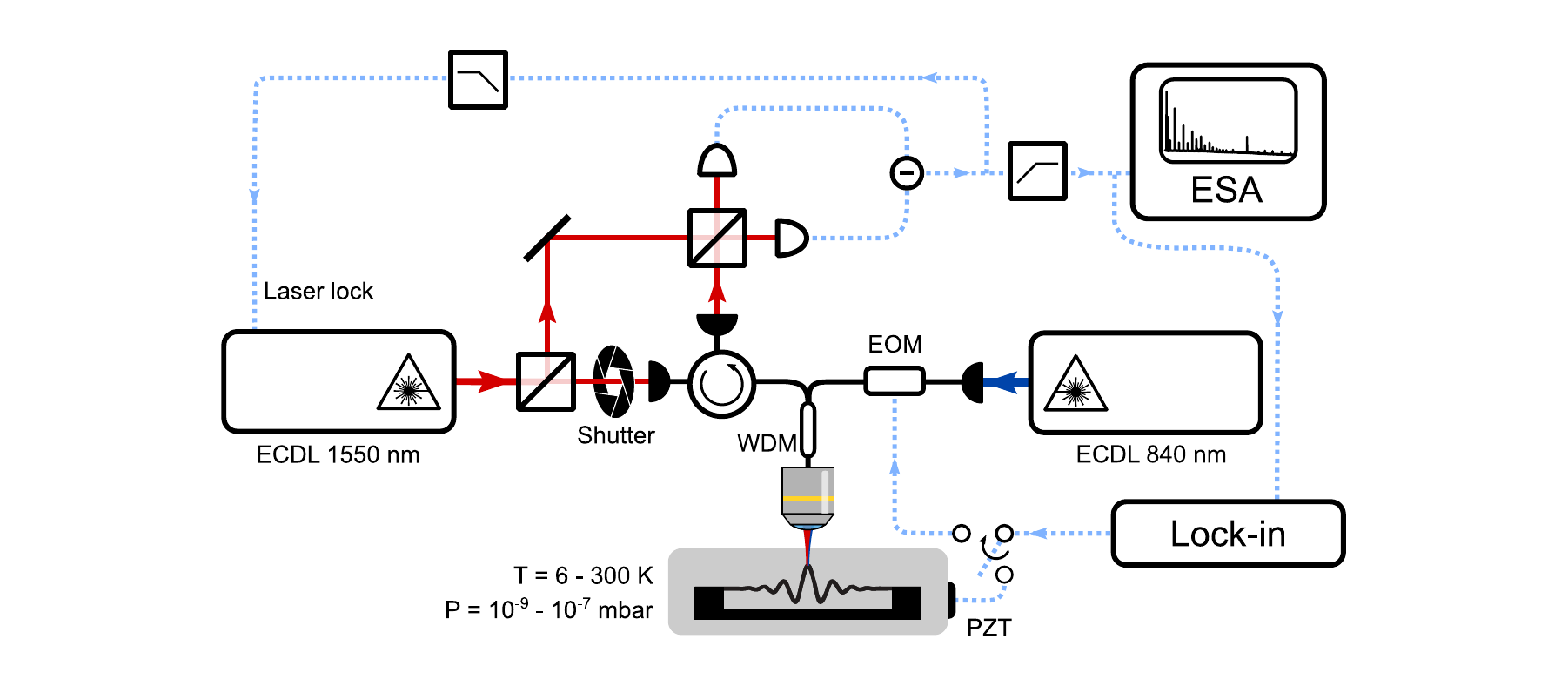}
\caption{\textbf{Schematic of the measurement setup.} ECDL: external cavity diode laser. EOM: electro-optical modulator. WDM: fiber optic wavelength division multiplexer. PZT: piezoelectric transducer. ESA: spectrum analyzer.
\label{fig:setup}}
\end{figure*}

We conduct the mechanical characterization measurements presented in the main text in a balanced Mach-Zehnder homodyne interferometer, in which we inject the light of a \SI{1550}{nm} tunable external-cavity diode laser (see \figref{fig:setup}). We mount the samples in a low vibration, closed-cycle Helium cryostat operating at temperatures as low as \SI{6}{K}. At the base temperature, the pressure inside the cryogenic vacuum chamber is lower than $\SI[retain-unity-mantissa = false]{1e-8}{mbar}$.

\SI{1550}{nm} light is focused on the string samples with a microscope objective optimized for near infrared transmission. We could finely adjust the position of the microscope column by means of a mechanical 3-axis stage. Part of the light reflected from the nanostring is collected by the objective and separated from the incident beam with a fiber optic circulator. A signal proportional to mechanical displacement is obtained by overlapping the reflected beam with an intense local oscillator ($P\approx \SI{1}{}-\SI{5}{mW}$) derived from the same laser; the interference pattern is locked to the optimal quadrature by actuating the laser frequency. The interference signal is converted to RF by a pair of balanced photodetectors. The interferometer can resolve thermomechanical noise down to the base temperature of the cryocooler.

In order to perform ringdown measurements, we excite the string samples either mechanically, with a piezoelectric actuator fixed on the sample mount, or photothermally, by sending an amplitude-modulated laser beam at \SI{840}{nm} through the microscope objective. We use a mechanical shutter to periodically suppress the probe laser and acquire gated ringdown traces.

\subsection*{Gas damping of the localized mode at room temperature}

\label{sec:gasdamp}

\begin{figure}[t]
\centering
\includegraphics[width=\columnwidth]{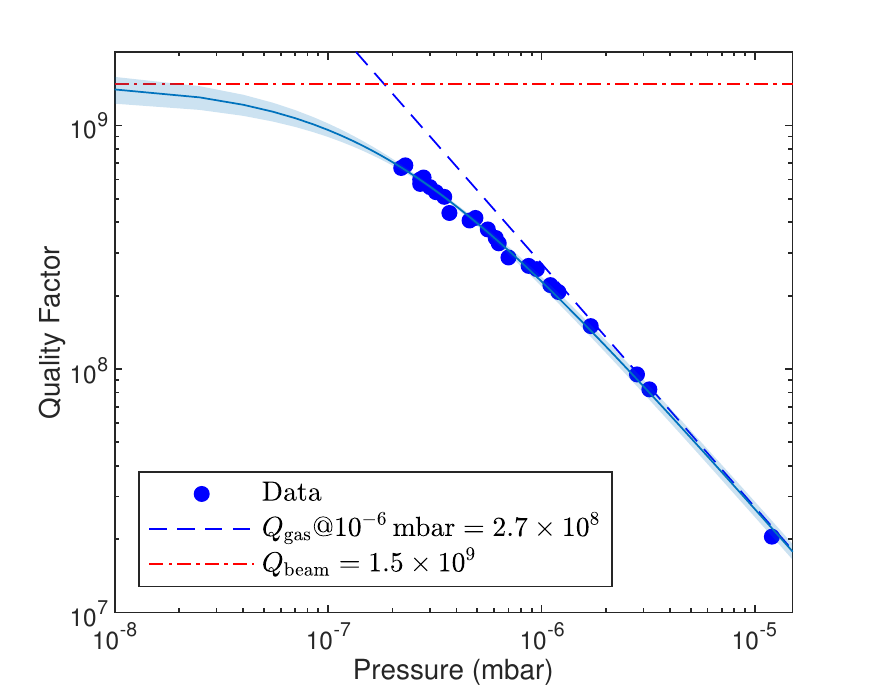}
\caption{\textbf{Quality factor as a function of pressure.} Solid line shows fit as described in the text, with 95\% confidence intervals shown by the shaded area.
\label{fig:gasdamping}}
\end{figure}

In the molecular regime, i.e. when the mean free path of the gas particles is much longer than the size of the structure, it can be found that the quality factor of a mechanical resonator will experience an extra damping rate due to random collisions with gas molecules, given by \cite{bao2002energy}:

\begin{equation}
Q_\mathrm{gas}=\frac{\rho\Omega_m h}{\Gamma_g p}
\end{equation}

\noindent where $p$ is the gas pressure, $\rho$ is the material density, $h$ is the film thickness and $\Gamma_g\propto\sqrt{m/2k_BT}$, where $m$ is the molecular mass of the gas and $T$ is the gas temperature. 

The sample described in section \ref{sec:QvsT} was partially gas damped around room temperature. At room temperature, the lowest attainable pressure in the vacuum chamber was around $p=\SI{2.2E-7}{\milli\bar}$, where we can estimate $Q_{\mathrm{gas}}\approx1.66\cdot10^9$. At this pressure, we found $Q=0.67\cdot10^9$ and therefore concluded our $Q$ was partially influenced by gas damping. We proceeded to measure the quality factor as a function of pressure in the chamber and then used a fit of form 

\begin{equation}
Q=\frac{1}{1/Q_{\mathrm{beam}}+1/Q_{\mathrm{gas}}}
\end{equation}

\noindent to extract the  quality factor of the beam in the absence of gas damping, $Q_\mathrm{beam}$ (see \figref{fig:gasdamping}).

At lower temperatures, the contribution of the gas damping rate was found to be negligible, due to significant cryopumping in the vacuum chamber.

\subsection*{Quality factors of additional samples}

\begin{figure}[t]
\centering
\includegraphics[width=\columnwidth]{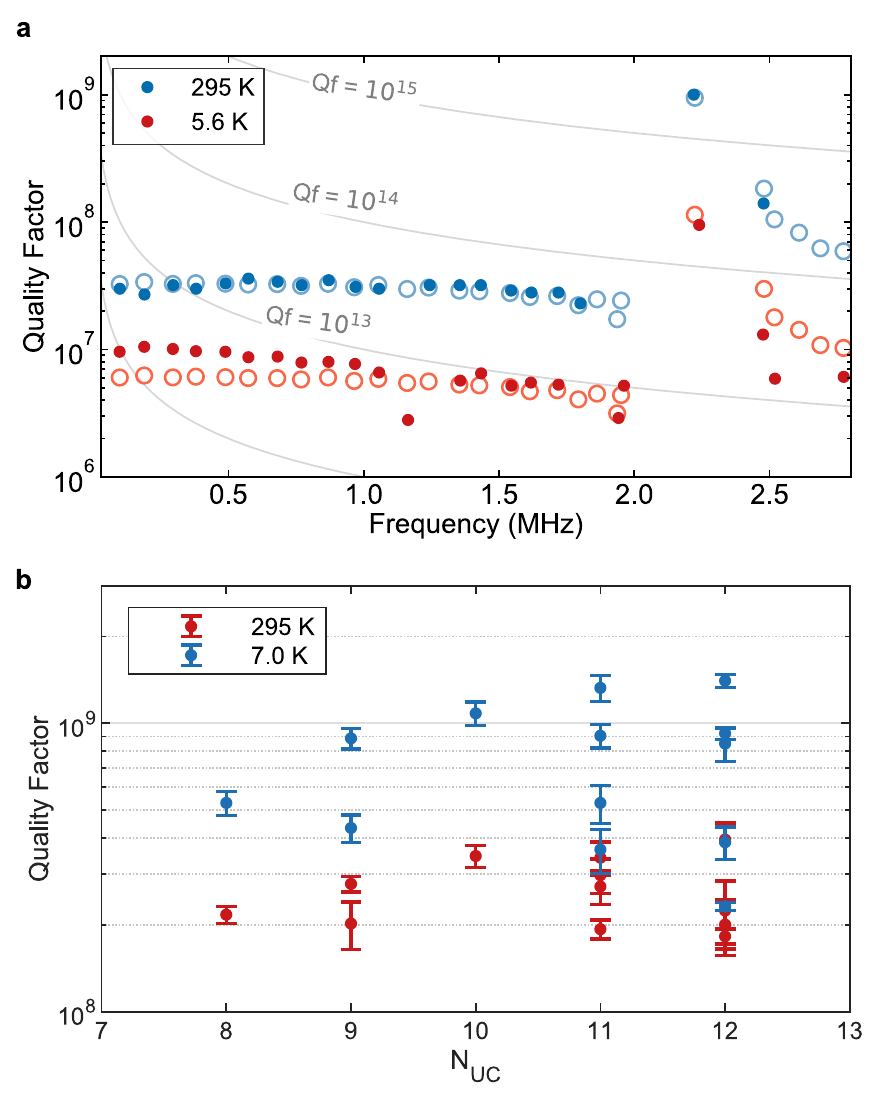}
\caption{\textbf{Quality factors of additional samples.} \textbf{a}, Quality factors of multiple mechanical resonances of a \SI{3.2}{mm}-long PnC string with 10 unit cells per side. Open circles represent a numerical model accounting for dissipation dilution via FEM. \textbf{b}, Quality factor versus number of unit cells per side of the defect, for a set of \SI{4.2}{mm}-long strings. The devices were characterized at room temperature and at the cryostat base temperature; different temperatures are distinguished by the colors of the dots. Error bars represent 95\% confidence intervals on Q, estimated from repeated measurements.
\label{fig:shortbeams}}
\end{figure}

We fabricated multiple chips with \SI{6.0}{mm}-long PnC strings, each containing 25 devices with a variable number of unit cells. A large variance of quality factors was observed, indicating probable contamination issues arising from the fabrication process. Among about $20$ measured devices, we observed a second string with the same design as the one presented in section \ref{sec:QvsT} and comparable mechanical parameters: $\Omega_m/2\pi = \SI{1.45}{MHz}$ and $Q \approx 6.4\cdot10^8$ at room temperature (limited by gas damping, as discussed in the previous section), and $\approx 6.3\cdot10^9$ at \SI{6.7}{K}.

Shorter devices exhibited generally a higher dissipation, in agreement with \eqref{eqn:DQ_SC}. In figure \figref{fig:shortbeams}a, we present a similar measurement of the quality factors for multiple mechanical modes of a \SI{3.2}{mm}-long PnC string, with 10 unit cells on each side of the defect. The characterization was repeated at room temperature and at \SI{5.6}{K}, and the data was fitted by a model accounting for dissipation dilution. For the cryogenic measurement, as in \figref{fig:SC}f, an additional boundary loss contribution was included to improve the agreement of model and data, allowing the inference of a silicon $Q_\mathrm{int} \approx 1.4\cdot 10^3$, significantly lower than for the \SI{6}{mm} device.

A batch of \SI{4.2}{mm}-long PnC strings showed a good yield of devices with high-Q localized modes, with quality factors between $10^8$ and $4\cdot10^8$ at room temperature. However, the improvement at liquid Helium temperatures was lower than for previously-discussed devices, suggesting a different dominant dissipation mechanism.
The number of unit cells was varied for these samples as well; by keeping the overall string length constant, we could localize mechanical modes between \SIrange{1.3}{2.1}{MHz}. No strong correlation between the number of unit cells and the Q of the localized mode was observed, as displayed in \figref{fig:shortbeams}b.

\subsection*{Estimating nanostring heating by its resonant frequencies}

\begin{figure*}[!t]
\centering
\includegraphics[width=\textwidth]{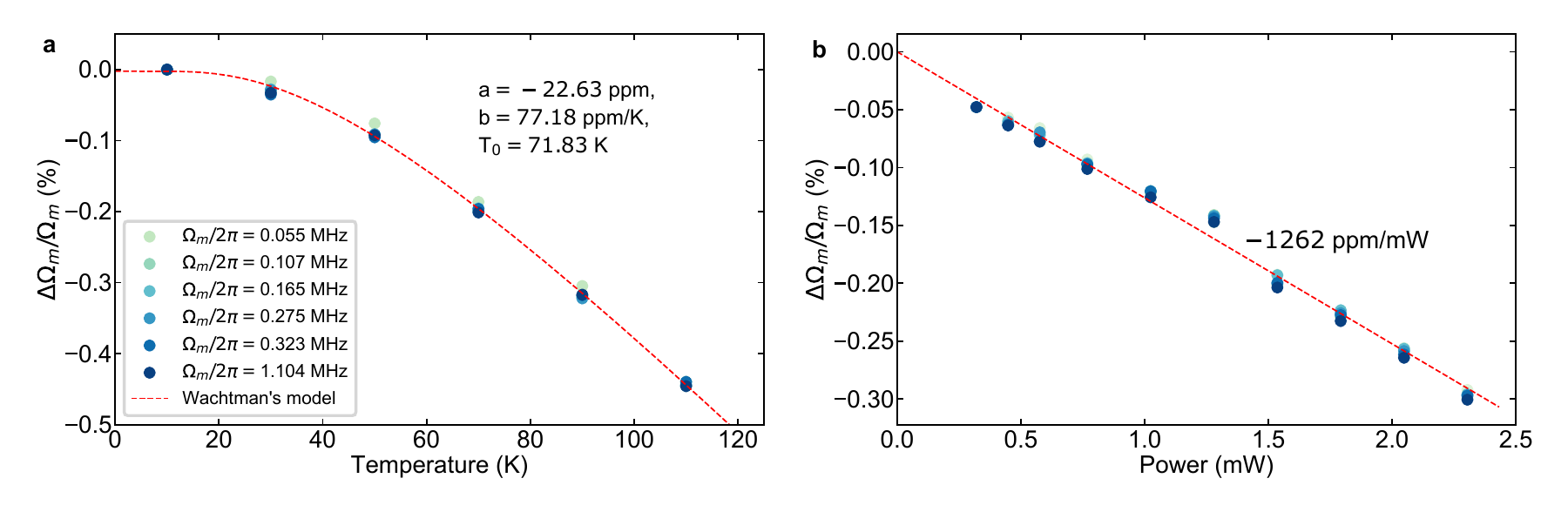}
\caption{\textbf{Response of resonance frequencies to temperature.} \textbf{a}, Relative variation of resonance frequencies, of multiple modes with the heater temperature, referred to the maximum measured frequency. The red dashed line is a fit to the curve $a - bT\mathrm{exp}\left(-T_0/T\right)$ \cite{anderson1966derivation}. \textbf{b}, Relative variation with variable impinging optical power, at $T = {10}{K}$. The red dashed line is a linear fit.
\label{fig:heating}}
\end{figure*}

The eigenfrequencies of the nanostrings can be computed with a perturbative approach \cite{younis2011mems}, starting from the solutions $u$ of the Euler-Bernoulli equation:

\begin{equation}
    \frac{d^2}{dx^2}\left(E I(x) \frac{d^2u}{dx^2}\right) - \mathcal{T}\frac{d^2u}{dx^2} + \rho_l(x) \frac{d^2u}{dt^2} = 0,
\end{equation}

\noindent where $u$ is the displacement pattern of flexural modes, $I(x) = w(x)h^3/12$ is the bending stiffness, $w(x)$ is the nanobeam width, $\mathcal{T}$ is the static tension force (constant along the nanobeam profile), $\rho_l = \rho w(x) h$ is the linear mass density, $\rho$ is the volume density of silicon, and $x$ is the string longitudinal coordinate, running from 0 to $L$.

The equation can be rewritten in terms of a normalized coordinate $s = x/L$, running from 0 to 1:

\begin{multline}
\lambda^2\sigma_\mathrm{avg}\frac{d^2}{ds^2}\left(\epsilon(s) v(s) \frac{d^2u}{ds^2}\right) -\\ 
- \sigma_\mathrm{avg}\frac{d^2u}{ds^2} + \rho(s)v(s)L^2 \frac{d^2u}{dt^2} = 0,
\end{multline}

\noindent where $\sigma_\mathrm{avg} = \mathcal{T}/\left(w_0h\right)$ is the stress along $x$ averaged over the nanostring, $w_0 = \frac{1}{L}\int^L_0 w(x)dx$ is the average string width, $v(s) = w(s)/w_0$ is a normalized width profile, and a potential position dependence was made explicit in the density and Young's modulus, $\rho = \rho(s)$ and $E = E_0\cdot\epsilon(s)$ . $\lambda = \sqrt{\frac{E_0}{12\sigma_\mathrm{avg}}}\frac{h}{L}$ as defined in the main text.

We can then separate $u$ in a product of space- and time- dependent parts, $u(s,t) = U(s)\cdot q(t)$, and write an harmonic oscillator equation for $q$ by multiplying the previous equation by $U(s)$ and integrating over $s$, using the clamped boundary conditions $U(0) = U(1) = 0$ and $U'(0) = U'(1) = 0$.  The spring constant $k$ of the oscillator is subdivided in two series components, $k^{(I)}$, associated with the bending stiffness, and $k^{(T)}$, related to the tension in the string. The spring constant and effective mass $m_\mathrm{eff}$ of the oscillator are:

\begin{align}
k^{\left(I\right)} &= \frac{\lambda^2\sigma_\mathrm{avg} w_0 h}{L}\int^1_0 \epsilon(s) v(s) \left(U''(s)\right)^2 ds \\
k^{\left(T\right)} &= \frac{\sigma_\mathrm{avg} w_0 h}{L}\int^1_0 \left(U'(s)\right)^2 ds \\
m_\mathrm{eff} &= h w_0 L \int_0^1 \rho(s) v(s) U^2(s) ds \\
\end{align}

Since $\lambda \ll 1$ for high aspect ratio strings under tension, we can neglect $k^{\left(I\right)}$ in the following steps. Eigenfrequencies are obtained as $\Omega_m \approx \sqrt{k^{\left(T\right)}/m_\mathrm{eff}}$.

We consider now the effect of a small variation in the physical properties of the string due to a non-uniform temperature profile $\theta(s)$ \cite{aguilarsandoval2015resonance}. Small relative variations in $\Omega_m$ are given by:

\begin{align}
    \frac{\Delta\Omega_m}{\Omega_m} &\approx \frac{\Delta k^{\left(T\right)}}{2k^{\left(T\right)}} - \frac{\Delta m_\mathrm{eff}}{2m_\mathrm{eff}} \\
    &= \frac{\Delta \sigma_\mathrm{avg}}{2\sigma_\mathrm{avg}} - \frac{\Delta L}{L} - \frac{\int_0^1 \Delta\rho(\theta(s)) v(s) U^2(s) ds}{2\cdot\int_0^1 \rho(s) v(s) U^2(s) ds} \label{eqn:df/f}
\end{align}

\noindent Note that an explicit dependence on the mode profile $U(s)$ is retained only through the variation of mass density (generally due to thermal expansion).

Using \eqref{eqn:df/f}, we interpret now the variation of resonant frequencies of a \SI{6}{mm}-long PnC nanostring with temperature, obtained by resolving thermomechanical peaks with a spectrum analyzer (see \figref{fig:setup}). The temperature was tuned from \SIrange{10}{110}{K} with a resistive heater fixed on the cryogenic mount. We notice in \figref{fig:heating}a that the relative frequencies decrease at a rate around \SI{-77}{ppm/K} beyond $\sim \SI{40}{K}$, and saturate at lower temperatures. The trend is similar to the predicted variation of Young's modulus at low temperatures \cite{anderson1966derivation,gysin2004temperature}, but it could also indicate a poor thermalization of the string to the sample plate at the lowest temperatures. Moreover, the magnitude of $\Delta\Omega_m/\Omega_m$ is more than one order of magnitude larger than the thermal expansion coefficient and has the opposite sign; hence we conclude that the stress variation (due to the change of Young's modulus or thermal strain) is the dominant effect in \eqref{eqn:df/f} in our temperature range. The presence of native oxide on the exposed surfaces of the string, however, may also influence the observed variation.

Absorption of \SI{1550}{nm} light has a similar effect on the string resonances, that drift towards lower frequencies as the optical power is increased. This effect is displayed in \figref{fig:heating}b, where the sample mount temperature is kept at \SI{10}{K} and power is measured at the microscope objective output (see \figref{fig:setup}). Note that in \figref{fig:heating}a, we accounted for the power dependence by recording at each temperature the value of the resonant frequencies as the optical power was varied, and extrapolating to vanishing power. As before, we do not observe any modeshape dependence of the relative frequency variation, suggesting that the last term in \eqref{eqn:df/f} does not relevantly contribute. The temperature profile in the nanostring is now expected to be strongly peaked at the laser position, due to strong variation of the conductivity with temperature.

We conclude therefore that absorption of \SI{1550}{nm} light, through two-photon processes or defect states, heats up significantly the \ch{Si} nanostring. However, $\Delta\Omega_m/\Omega_m$ in \figref{fig:heating}a and b cannot be directly compared, as when optical power is absorbed locally, a large temperature difference is established between the string and the substrate chip, probably leading to stronger variations of $\sigma_\mathrm{avg}$. 

When the laser power is stepped or the shutter is opened, resonant frequencies are observed to relax at characteristic times below \SI{1}{s}. This justifies the use of the gated ringdown technique to characterize mechanical damping, and the assumption that the string temperature is close to the temperature of the cryostat.

\bibliography{references}

\section*{Data availability}

Data supporting the manuscript figures and lithographic masks used for microfabrication will be released on \texttt{Zenodo} upon publication of this work.

\section*{Acknowledgements}

The authors thank Dalziel Wilson and Amir Ghadimi for their contributions in the early stages of the project, and Robin Groth for assistance with sample characterization. This work was supported by funding from the Swiss National Science Foundation under grant agreement no. 182103, the EU H2020 research and innovation programme under grant agreement no.732894 (HOT) and the European Research Council grant no. 835329 (ExCOM-cCEO). N.J.E. acknowledges support from the Swiss National Science Foundation under grant no. 185870 (Ambizione). This work was further supported by the Defense Advanced Research Projects Agency (DARPA), Defense Sciences Office (DSO) contract no. HR00111810003. All samples were fabricated at the Center of MicroNanoTechnology (CMi) at EPFL.

\section*{Author contributions}
A.B. fabricated the devices with assistance from M.J.B. The devices were characterized by A.B., D.A.V, and N.J.E. The characterization setup was constructed by S.A.F., N.J.E and A.B. The transmission electron microscopy was performed by V.B. The manuscript was written by A.B. and N.J.E. with assistance from the other authors. The work was supervised by N.J.E and T.J.K.
\end{document}